\newcommand{\beq}{\begin{equation}}
\newcommand{\eeq}{\end{equation}}
\newcommand{\beqn}{\begin{eqnarray}}
\newcommand{\eeqn}{\end{eqnarray}}
\newcommand{\nn}{\nonumber}

\documentclass[12pt]{article}
\begin{document}

\title{{\bf A 0-dimensional counter-example\\ to rooting?}}

\author{{\large{G.C. Rossi}}$^{^{a)b)*)}}$
{\large{M. Testa}}$^{^{c)d)**)}}$\\\\
\small $^{a)}$Dipartimento di Fisica, Universit\`a di Roma {\it Tor Vergata}\\
\small $^{b)}$INFN, Sezione di Roma 2 \\
\small Via della Ricerca Scientifica 1, 00133 Roma, Italy\\
\small ${^{c)}}$Dipartimento di Fisica, Universit\`a di Roma {\it La Sapienza}\\
\small $^{d)}$INFN, Sezione di Roma {\it La Sapienza}\\
\small P.le A. Moro 2, 00185 Roma, Italy} 
\date{}
\maketitle
\begin{abstract}
We provide an example of a 0-dimensional field theory where rooting 
does not work.
\end{abstract}

\vskip 4.cm
\noindent\small {$^{* )}$ E-mail address: giancarlo.rossi@roma2.infn.it}\\
\noindent\small {$^{* *)}$ {\it Correspondence to:} M. Testa, Phone:
+39-0649914289. \\E-mail address: massimo.testa@roma1.infn.it}\\
\vfill

\newpage

Rooting of staggered fermions~\cite{Kogut:1974ag,MILCH} is a necessary step to 
yield the correct number of degrees of freedom circulating in loops in lattice QCD 
simulations with Kogut--Susskind fermions. Though rooting appears correct in perturbation 
theory, it has been questioned on the basis of either problems with locality~\cite{LOC} or for a 
possibly incorrect chiral behaviour~\cite{CRE}. Both criticisms have been argued 
not to apply to the calculation of physical observables in refs.~\cite{SHA} 
and~\cite{SETAL} (see also~\cite{REV}).


In this short note we would like to present an oversimplified field theoretical 
model in zero dimensions hinting at serious difficulties with rooting. Consider 
the (0-dimensional euclidean functional) integral with two tastes 
\beq
Z^{(2)}(j)=\int_{-\infty}^{+\infty} dx \,{\rm e}^{-\frac{x^2}{2}+jx} 
\int \prod_{i=1,2}d\bar \psi_i d\psi_i 
\exp{\left(\sum_{i=1,2}\bar\psi_i(gx+m)\psi_i \right)}\, ,
\label{Z2}
\eeq
where $x$ is a $c$-number and $\bar \psi_1\, , \psi_1\, ,\bar \psi_2\, ,\psi_2$ are 
Grassmann variables. The parameters $m>0$ and $g$ are real numbers with the role 
of mass and coupling constant, respectively. To mimic field theory we have also introduced 
an external source coupled to the ``bosonic field'', $x$. Performing the fermionic integration 
in~(\ref{Z2}), one immediately gets
\beq
Z^{(2)}(j)=\int_{-\infty}^{+\infty} dx \,
{\rm e}^{-\frac{x^2}{2}+jx} (gx+m)^2= 
\sqrt{2\pi}\,{\rm e}^{\frac{j^2}{2}}\left[g^2+(gj+m)^2\right]\, .
\label{Z2F}
\eeq
One notices that, as expected, the result of the two-taste fermion integration extended 
over a finite volume (in this case just one site) is the square of a polynomial in the fields. 
Thus in principle there is no problem in taking the square root of the fermionic determinant.

We now want to expose the difference between the direct computation of the ``partition 
function'' with a single taste, $Z^{(1)}$, and the evaluation where the rooting trick is 
exploited. The direct calculation of $Z^{(1)}$ gives 
\beqn
&&Z^{(1)}(j)=\int_{-\infty}^{+\infty} dx \,{\rm e}^{-\frac{x^2}{2}+jx} 
\int d\bar\psi d\psi\exp{\left(\bar\psi(gx+m)\psi\right)}=\nn\\
&&\phantom{Z^{(1)}(j)}
=\int_{-\infty}^{+\infty} dx\,{\rm e}^{-\frac{x^2}{2}+jx} (gx+m) 
= \sqrt{2\pi}\,{\rm e}^{\frac{j^2}{2}}\left[gj+ m\right] \, ,
\label{Z1}
\eeqn
while using rooting, one obtains 
\beqn
&&Z^{(1)}_{\rm root}(j)=\int_{-\infty}^{+\infty} dx\, 
{\rm e}^{-\frac{x^2}{2}+jx} \left[\int \prod_{i=1,2}d\bar \psi_i d\psi_i 
\exp{\left(\sum_{i=1,2}\bar\psi_i(gx+m)\psi_i \right)}\right]^{1/2}=\nn\\
&&\phantom{Z^{(1)}_{\rm root}(j)}
=\int_{-\infty}^{+\infty} dx \,{\rm e}^{-\frac{x^2}{2}+jx} |gx+m|=\nn\\
&&\phantom{Z^{(1)}_{\rm root}(j)}=2\,{\rm e}^{\frac{j^2}{2}} \left[g\,{\rm e}^{-\frac{{(gj+m)}^2}{2g^2}} 
+[gj+m]\int_0^{\frac{gj+m}{g}} d\eta\,{\rm e}^{-\frac{\eta^2}{2}} \right]\, .
\label{Z2R}
\eeqn
The difference between eqs.~(\ref{Z1}) and (\ref{Z2R}) vanishes exponentially for small $g$. 
This tells us that, as expected, $Z^{(1)}(j)$ and $Z^{(1)}_{\rm root}(j)$ have an 
identical perturbative expansion, but drastically differ non perturbatively. 

The origin of this non perturbatively different behaviour is to be traced back, in our example, 
to the existence of zero-mode configurations around which the regularized determinant does not 
necessarily have a fixed sign. This shows that the real point in order to settle the question of the 
validity of rooting is to decide if the regularized staggered determinant is going to vanish 
somewhere in gauge field space. More generally, since there is no regularized functional 
integral corresponding to the rooting prescription, it should be investigated what sort of 
modifications would be required at large, non perturbative gauge configurations for the 
rooting prescription to lead to an acceptable Quantum Field Theory.


\vspace{.2cm}
{\bf Acknowledgments -} We wish to thank S. Sharpe for comments and R. Frezzotti for discussions.

\end{document}